# Geometrical Field Theory of Hamilton Dynamic System In Rational Mechanics

Xiao Jianhua

Measurement Institute, Henan Polytechnic University, Jiaozuo, China

**Abstract:** When a set of particles are moving in a potential field, two aspects are concerned: 1) the relative motion of particle in spatial domain; 2) the particle velocity variations in time domain. The difficulty on treating the systems is originated from the fact that the motion in time domain and the motion in spatial domain are coupled together completely. Generally, for a Hamilton dynamic system established by a set of general velocity functions, several abstract theories have been well established, such as Lie algebra, Symplectic manifold, Poisson brackets, and others. However, mathematically, to find out a general Hamilton function is very difficult even for very simple problems. Inspired by these abstract mathematic researches, the Hamilton dynamic system is studied by geometrical field theory of deformation. Firstly, referring to the instant configuration, the deformation tensor in spatial domain and the velocity transformation tensor in time domain are established for a dynamic system defined by a set of general velocity functions. Secondly, the general deformation tensor in velocity space domain is obtained. From continuum mechanics point, the stress tensor is defined through introducing background feature of space. Then, the general motion equations are established. Based on them, the simple motions are divided into two classes: 1) stable motion; and 2) radiating motion. Both of them have quantum solution structures. The features of space-time continuum are studied to obtain some intrinsic understanding about basic physical facts. This research shows an engineering way to treat the Hamilton dynamic system.

Contents

## 1. Introduction

Traditional Hamilton dynamic system which describes the particle moving in pure potential field (independent of velocity and time) is studied. The free mass point motion in gravity field is a typical example. For Hamilton dynamic system [1], Lie algebra is its general geometrical features. By Lie algebra formulation, the general mathematic structures are established. Unfortunately, this is achieved by losing the apparent meaning of physical quantities. For theoretic researches, this is acceptable. But, for engineering application, this indeed is a big problem. Looking for an engineering way to handle Hamilton dynamic system is the topic for this research.

In this paper, the exterior particle moving in continuum is studied. This topic has been treated repeatedly in different science branches. This research will formulate a geometrical field theory to generalize the related researches based on rational mechanics view-point.

The most significant feature of a general Hamilton system is that: the time derivatives of spatial position vector and the time derivatives of moment vector have the Lie algebra structure [2]. However, the related proves are mainly based on physical motion equations where the motion has no effects on potential field is implied. Particle motion in potential field is its traditional application field.

In fact, the Hamilton quantity is defined as the addition of kinetic energy (purely determined by moment vector field) and potential field (purely determined by position). Logically, the kinetic energy has no coupling with the potential field. Their coupling is determined by the Hamilton dynamic system equations. In conventional time-space continuum, this is highly acceptable. However, when the continuum is composed by dense material distribution, the potential field should include the contribution of medium deformation which is caused by the motion of material particles moving in the medium. There are many practical problems can be attributed by this phenomenon.

Although continuum deformation mechanics is well established, the exterior particles moving in continuum still is a difficult problem. The main features are that: the exterior particles will cause the global deformation of continuum [3]. As a result, the potential field will depend on the exterior particle moving velocity. Hence, the potential field in Hamilton formulation is moment and position dependent.

Then, in abstract sense, the interaction between particle motion and potential field must be included. As a direct instinct understanding, the Lie algebra should be reexamined for its applicability. Furthermore, the potential field produced by particles themselves make the interaction among particles must be included. The late usually is referred as many body problems.

What physical philosophy is underling Hamilton formulations? This is the kernel problem in this research. Once the physical consideration is exposed, an alternative way is constructed. This is the main topic for this paper. Through out the paper, the thoughts of general relativity are implied. That is to say, the mass and field interaction is under consideration.

Summing above considerations, the problem is formulated as the following. When a set of particles are moving in a potential field, two aspects are concerned: 1) the relative motion among particles in spatial domain; 2) the particle velocity variations in time domain. The difficulty on treating the so-called many body systems is originated from the fact that the motion in time domain and the motion in spatial domain are coupled together completely. Generally, for a Hamilton dynamic system established by a set of general velocity functions, several abstract theories have been well established, such as Lie algebra, Symplectic manifold, Poisson brackets, and others. Inspired by these abstract mathematic researches, the Hamilton dynamic system is studied by geometrical field theory of deformation.

For this purpose, the Hamilton system is reviewed firstly. Then, the problem is explained as a general deformation of manifold. Based on this new understanding, the deformation tensor is applied to a general classical dynamic system to form the related geometrical field theory. Finally, a geometrical field theory of Hamilton dynamic system is established. To show its value, the quantum mechanics is obtained as the natural results of such a formulation applied to basic particles. In this sense, this research shows a way from classical physics to quantum mechanics in logic consistent form. It is hopped that this research will promote the application of geometrical field based on deformation concept to dynamic system.

## 2. Simple Review of Classical Formulation of Hamilton Dynamic System

When the potential field only depends on one set of dynamic quantities (say position coordinator), the geometrical field of Hamilton dynamic system is well formulated. They are the starting points in this research. Hence, a simple review may be helpful for understanding this paper.

### 2.1 Classical Dynamic System

For a classical dynamic system defined by the velocity vector $\frac{dx^i}{dt}\vec{g}_i$, $i=1,2,3$ the conventional motion equation is defined as:

$$\frac{dx^i}{dt} = f^i(\eta, x) \tag{1}$$

Where, the $\eta$ is a set of physical parameters. The system is determined by initial conditions: $x^i(t,x)\big|_{t=0}$ are known velocity field. The particle motion is caused by potential field, which only depends on position and is independent to time parameter. In modern physics, as the inertial force is interpreted as the results of potential field [4], the above dynamic system is very general.

In practical engineering problems, the function $f^i(\eta,x)$ is viewed as known field. The position field is wanted.

*2.2 Hamilton Equations*

Through introducing the moment vector $p_i$ and Hamilton function $H(\eta,x,p)$, the Hamilton principle gives out the general motion equation as canonical form:

$$\frac{dp_i}{dt} = -\frac{\partial H}{\partial x^i} \tag{2-1}$$

$$\frac{dx^i}{dt} = \frac{\partial H}{\partial p_i} \tag{2-2}$$

When each of variables sets $\{x\}$ and $\{p\}$ are independent.

In fact, for Equation (1), taking time derivative for both sides, one obtained that:

$$\frac{dv^i}{dt} = \frac{d^2 x^i}{dt^2} = \frac{\partial f^i}{\partial x^j} \cdot \frac{dx^j}{dt} = v^j \cdot \frac{\partial f^i}{\partial x^j} \tag{3}$$

Where (and here after), the repeat index shows summation. Then, letting $\rho_{ij} = \rho_{ji}$ be the mass density tensor (physical constants), and defining $p_i = \rho_{ij} v^j$, the Equation (3) can be re-expressed as:

$$\frac{dp_i}{dt} = \rho_{ij} v^l \frac{\partial f^j}{\partial x^l} = \rho_{ij} f^l \frac{\partial f^j}{\partial x^l} \tag{4}$$

By comparing it with Equation (2-1) and comparing Equation (1) with Equation (2-2), one obtains:

$$\frac{\partial H}{\partial p_i} = f^i, \quad \frac{\partial H}{\partial x^i} = -\rho_{ij} f^l \frac{\partial f^j}{\partial x^l} \tag{5}$$

Hence, the velocity vector function is related with a pure scalar function. Generally speaking, through formulating:

$$[\frac{\partial H}{\partial p_i}, \frac{\partial H}{\partial x^j}] = \frac{\partial H}{\partial p_i}\frac{\partial H}{\partial x^j} - \frac{\partial H}{\partial p_j}\frac{\partial H}{\partial x^i} = f^i \cdot (\rho_{jk} f^l \frac{\partial f^k}{\partial x^l}) - (\rho_{ik} f^l \frac{\partial f^k}{\partial x^l}) \cdot f^j \tag{6}$$

Rewritten it as the abstract form:

$$[\frac{\partial H}{\partial p_i}, \frac{\partial H}{\partial x^j}] = (f^i \cdot \rho_{jk} - \rho_{ik} \cdot f^j) \cdot \frac{\partial f^k}{\partial x^l} f^l = C_l^{ij} \cdot f^l \tag{7-1}$$

Where,

$$(f^i \cdot \rho_{jk} - \rho_{ik} \cdot f^j) \cdot \frac{\partial f^k}{\partial x^l} = C_l^{ij} \tag{7-2}$$

Or, in the form:

$$[\frac{\partial H}{\partial p_i}, \frac{\partial H}{\partial x^j}] = (f^i \cdot \rho_{jk} - \rho_{ik} \cdot f^j) \cdot \frac{\partial f^k}{\partial x^l} f^l = \tilde{C}_l^{ij} \cdot \tilde{f}^l \tag{8-1}$$

Where,

$$(f^i \cdot \rho_{jk} - \rho_{ik} \cdot f^j) = \tilde{C}_l^{ij}, \quad \frac{\partial f^k}{\partial x^l} f^l = \tilde{f}^l \qquad (8\text{-}2)$$

Not that: $C_k^{ij} = -C_k^{ji}$, and $\tilde{C}_k^{ij} = -\tilde{C}_k^{ji}$.

Therefore, by suitably defining a base vector field $\vec{V}_1, \vec{V}_2, ..., \vec{V}_n$ on manifold $M^n$, if they meet the relation equation:

$$[\vec{V}_i, \vec{V}_j] = c_{ij}^k \vec{V}_k \qquad (9)$$

Then, Lie algebra formulation can be established for Hamilton dynamic system. Here, the main point of the research is that: 1) for single particle, the mass density tensor is $\rho_{ij} = \rho \delta_{ij}$; 2) for continuum, the mass density tensor is determined by potential field. As an example, in general relativity $\rho_{ij} \to g_{ij}$, the mass density tensor plays the role of gauge field [5].

Hence, the conclusion is that: a geometrical gauge tensor (determined by matter in motion) is embedded in the Lie algebra structure. As a natural result, recovering the implied geometrical gauge tensor will lead to another formulation of Hamilton dynamic system. This is the target of this research.

### 2.3 Exterior Differential 2-form

As a general extension, through define general coordinator $q_i$, $i = 1, 2, ..., n$ and their complement momentum coordinator $p_i$, $i = 1, 2, ..., n$, the abstract algebra formulation of Hamilton dynamic system is obtained as:

$$\frac{dq_i}{dt} = [q_i, H] \qquad (10\text{-}1)$$

$$\frac{dp_i}{dt} = -[p_i, H] \qquad (10\text{-}2)$$

Where, Poisson bracket is defined as:

$$[F, G] = \frac{\partial F}{\partial q_i} \frac{\partial G}{\partial p_i} - \frac{\partial F}{\partial p_i} \frac{\partial G}{\partial q_i} \qquad (10\text{-}3)$$

Generally speaking, taking $H(p,q)$ as a function in manifold $M^{2n}$, the problem is formulated as the differentiable mapping in multiplicative space $T_x M \times T_x M$:

$$\omega^2(x) : T_x M \times T_x M \to R \qquad (11)$$

It has the algebra features: 1) bi-linearity;

$$\omega^2(x)(a\vec{V}_1 + b\vec{V}_2, \vec{V}_3) = a\omega^2(x)(\vec{V}_1, \vec{V}_3) + b\omega^2(x)(\vec{V}_2, \vec{V}_3) \qquad (12\text{-}1)$$

2) anti-symmetry;

$$\omega^2(x)(\vec{V}_1, \vec{V}_2) = -\omega^2(x)(\vec{V}_2, \vec{V}_1) \qquad (12\text{-}2)$$

3) closeness;

$$d\omega^2(x) = 0 \qquad (12\text{-}3)$$

4) non-singularity.

$$\omega^2(x)(\vec{V}_1, \vec{V}_2) \neq 0, \text{ for } \vec{V}_1 \neq 0 \tag{12-4}$$

Such a symplectic structure is already established in theoretic physics [6-7]. In mathematic sense, instead of looking for the trace solution $x(t)$ of Equation (1), the Hamilton equation (2) is looking for the general function $H(\eta, x, p)$ or $H(\eta, p, q)$. Once it is obtained, the position and velocity information are obtained for actual engineering problems.

On physical sense, the different dynamic system has different Hamilton function. Once the Hamilton function is formulated correctly, the system dynamic behavior is obtained.

Therefore, the Hamilton dynamic system becomes the common topic for mathematics and physics.

However, in many real engineering problems, the explicit form of $H(\eta, p, q)$ is unknown even we can construct it with very strong reasons. In contrast, the function $f^i(\eta, x)$ (instant velocity field) may be well measured. So, if we accept the general formulation of Hamilton dynamic system, the $[q_i, H] \, (= \frac{dq_i}{dt})$ is known, while the $[p_i, H] \, (= -\frac{dp_i}{dt})$, the acceleration force, is the unknown field, which we are looking for.

### 2.4 P. Dirac Form

To emphasize the half known $F \in C^\infty(M^n)$ and half unknown $G \in C^\infty(M^n)$ features of Hamilton function, through defining vector operators $\vec{V}_F \equiv [F, \bullet]$, $\vec{V}_G \equiv [G, \bullet]$, the 2-form on manifold $M^n$ is obtained as (P. Dirac form):

$$\omega^2(\vec{V}_G, \vec{V}_F) = [F, G] \tag{13}$$

### 2.5 Canonical Transformation

Traditionally, for coordinators $P, Q \in M^{2n}$ and $p, q \in M^{2n}$, the looking for unknown canonical transformation: $p, q \to P, Q$ under primitive function $S(p, q)$ is defined by the canonical equation:

$$PdQ - pdq = dS(p, q) \tag{14}$$

It has already proven that: 1) if for any path on the manifold can be contracted to a point, the path integral $\oint_\gamma PdQ = \oint_\gamma pdq$ equation is met, then the transformation $p, q \to P, Q$ is canonical.

2) if $[F, G]_{P,Q} = [F, G]_{p,q}$ is met for all $F, G \in C^\infty(M^n)$, then the transformation $p, q \to P, Q$ is canonical.

As a natural conclusion, the Hamilton equation form is invariant under canonical transformation:

$$\dot{F}_{P,Q} = \dot{F}_{p,q} = [F, H]_{p,q} = [F, H]_{P,Q} \tag{15}$$

This formulation is widely used in textbook.

*2.6 Geometrical transformation of Points Set*

No matter what mathematic system is adopted, the key physical problems [8-9] are: 1) if velocity field is measured as known field, the general force field is looking for; 2) if general force field is known field, the general position field is looking for. In abstract sense, the problem is bi-linear and anti-symmetric through the Hamilton formulation.

What physical philosophy is underling these formulations? This is the kernel problem in this research.

Generally reasoning on physical sense, the particle motion in potential field can be viewed through two main features [10-11]: 1) the particle moving trace is deformed by the force produced by the potential field in time domain (acceleration); 2) the particle moving trace is deformed by action from other particles in spatial domain (elastic deformation).

This reasoning is supported by conventional observations: 1) relative distance variation among particles; 2) moving velocity variation along its trace. Both of them determine the gauge tensor of the manifold.

The first phenomenon is described by spatial deformation tensor, which is used in continuum mechanics; the second phenomenon is described by the particle acceleration along its moving trace (Newton force). The continuum feature is completely determined by the potential field. The gravity field is implied by introducing the Newton force.

## 3. Geometrical Field Formulation of Hamilton Dynamic System

Considering a material element in continuum, as the continuum is composed by the same material elements, the material element is moving in the continuum itself-belonging. Hence, the potential field is completely determined by the continuum deformation. This is the basic view-point for continuum deformation mechanics. Or, say Hook mechanics view-point. In this case, the Hamilton quantity is the kinetic energy adds the deformation energy. For static problems, the deformation energy plays the essential role in continuum deformation mechanics. It works well.

For mass point moving in invariant potential field (time-independent), the classical Hamilton quantity is the addition of kinetic energy and the pre-settled potential field. The mass and the potential field have no intrinsic relations, although the mass point motion is controlled by both energy fields. For pure pre-existing potential field, if the mass points have no contribution to the potential field, the Hamilton mechanics works well. This, in essential sense, is based on Newton mechanics view-point.

However, for many body dynamic problems, when the Hamilton mechanics is applied, the solutions are very difficult to be obtained. Generally, the potential fields depend on the relative position of mass points (or say their configuration). As each mass is in dynamic moving state, the configuration is in dynamic evolution state. Hence, the configuration variation is controlled by the relative motion velocity fields among mass points. Physically, the position coordinator and the velocity field play different roles. In this case, as the configuration variation is controlled by the relative velocity field variation in spatial domain, the potential field should be the explicit function of position coordinator and relative velocity variation rather than the explicit function of position coordinator and velocity.

Hence, based on physical consideration, the potential field in Hamilton formulation should not be the explicit function of position coordinator and velocity (although many mathematics formulation taken it as apparent). This is the main problem for many body problems.

On the other side, for dynamic deformation of continuum (such as fluid flow), the kinetic energy may be explicitly position dependent, such as turbulent flow or vortex flow. As the Newton force explicit dependents on deformation tensor (defined by the relative velocity field variation in spatial domain), the kinetic energy becomes dependent on relative velocity among material elements. Navier-Stokes motion equation makes this point very clear.

Hence, the general form of Hamilton quantities should be: 1) the kinetic energy explicitly depends on velocity and velocity spatial-gradient; 2) the potential field explicitly depends on position and velocity spatial-gradient.

Based on above reasoning, the geometrical field description of Hamilton dynamic system is formulated. This is the main topic for this section.

*3.1 Manifold Deformation defined by Relative Flow*

For continuum, considering the classical dynamic system defined in three dimensional space ($x^i, i=1,2,3$) embedded in the continuum, for any material element, the Newton velocity field is given as:

$$\frac{dx^i}{dt} = f^i(\eta, x) \tag{16}$$

When the continuum is deformable, its geometrical field form is expressed as:

$$\frac{d\vec{x}}{dt} = f^i(\eta, x)\vec{g}_i \tag{17}$$

Where, $\vec{g}_i = \vec{g}_i(x)$ are the instant base vectors for unit coordinator along the coordinator $x^i$ direction. When the spatial continuum is Newton space, the inertial form in standard rectangular coordinator system, the continuum has no deformation. In this case, by definition, $g_{ij}^0 = \delta_{ij}$, the $\vec{g}_i^0 = \{\vec{i}, \vec{j}, \vec{k}\}$, as it is in usual standard textbook.

As the solution of Equation (16) gives out the motion path $x(x_0, t)$ from initial position $x_0$ to current position, the base vectors are determined by motion path also. Putting this in mine, the set of paths of the system forms an evolution manifold with time as its parameters. Focusing our attention on the spatial features, the space continuum is deformed. For any instant, the gauge tensor $g_{ij}(x)$ determines the configuration of manifold, which is defined as:

$$g_{ij} = \vec{g}_i \cdot \vec{g}_j \tag{18}$$

For mass points set, the relative path variation can be attributed as the deformation of configuration. To emphasis the physical aspects, the instant configuration at $x_0$ is taken as reference configuration, their base vectors are expressed $\vec{g}_i^0$. By definition, $g_{ij}^0 = \delta_{ij}$.

For unit time (suitably selected for the real problems), the instant deformation tensor [12] is defined as:

$$F_j^i(x) = \delta_j^i + \frac{\partial f^i}{\partial x^j} \tag{19}$$

By this formulation, the path instant variation (unit time variation) is expressed as:

$$\vec{g}_i = F_i^j \vec{g}_j^0 = (\delta_i^j + \frac{\partial f^j}{\partial x^i})\vec{g}_j^0 \tag{20}$$

The deformation tensor $F_i^j$ is defined at point $(t, x(t+1))$ referring to point $(t, x(t))$.

Comparing with conventional Hamilton formulation, $\frac{\partial f^i}{\partial x^j} = \frac{\partial H(\eta, x, p)}{\partial p_i \partial x^j}$. So, it is easy to identify that: for Hamilton dynamic system, the momentum variable is eliminated in this research. Mathematically, this way will simplify the solution process for some problems.

By this way, the path variation is attributed to the spatial deformation caused by global effects of dynamic system. In essential sense, this view-point is a copy of gravity field theory in modern physics. As an example in special relativity, the velocity field will cause local spatial deformation. For $\frac{d\vec{x}}{dt} = V\vec{g}_3^0$, special relativity gives out:

$$\vec{g}_1 = \vec{g}_1^0, \quad \vec{g}_2 = \vec{g}_2^0, \quad \vec{g}_3 = \frac{1}{\sqrt{1 - \left(\frac{V}{c}\right)^2}} \cdot \vec{g}_3^0 \tag{21}$$

Where, the $c$ is light velocity constant. It is identical with the spatial parts of Lorentz coordinator transformation. By comparing with our definition equation (19), it is found that:

$$\frac{\partial V}{\partial x^3} = \frac{1}{\sqrt{1 - \left(\frac{V}{c}\right)^2}} - 1, \quad \frac{\partial V}{\partial x^1} = 0, \quad \frac{\partial V}{\partial x^2} = 0 \tag{22}$$

In elastic deformation sense, the space in motion direction is stretching-out.

In classical fluid mechanics, taking the Eular coordinator system as the reference configuration, the strain rate which expresses the instant relative configuration variation is defined as:

$$\dot{\varepsilon}_{ij} = \frac{1}{2}(\frac{\partial f^i}{\partial x^j} + \frac{\partial f^j}{\partial x^i}) \tag{23}$$

It shows that the local instant deformation is determined by the velocity gradient.

Supported by above general reasoning, the spatial continuum deformation is defined by Equations (19) and (20). It clear that here the continuum is in very general sense.

*3.2 Manifold Deformation defined by Acceleration*

In Newton mechanics, the acceleration is related with inertial force. It is the force that makes the motion path changes. Putting this in mine, the acceleration along a path of the system is formulated as:

$$\frac{d^2\vec{x}}{dt^2} = f^j \cdot \frac{\partial f^i}{\partial x^j} \cdot \vec{g}_i \tag{24}$$

Therefore, for unit time incremental, the velocity variation field is expressed as:

$$f^i(\eta, x)\big|_{t+1} = [f^i(\eta, x) + f^j \frac{\partial f^i}{\partial x^j}]\big|_t \tag{25}$$

Hence, in phase space (velocity space), the instant deformation (unit time deformation) is defined as:

$$f^i = (\delta^i_j + \frac{\partial f^i}{\partial x^j}) f_0^j = F^i_j \cdot f_0^j \tag{26}$$

Where, the low index shows the reference initial time. The deformation tensor $F_i^{\,j}$ is defined at point $(t+1, x(t))$ referring to point $(t, x(t))$.

In fluid mechanics, taking the Eular coordinator system as the reference configuration, the velocity spatial variation caused acceleration is a well-known item in Navier-Stokes equation. So, the above formulation is based on strong background.

Note that, although the physical meanings of Equations (20) and (26) are different, the deformation tensor is calculated at point $(t, x(t))$.

### 3.3 Path Deformation Tensor on Material Motion Path in Velocity Space

Now, considering the instant variation (unit time variation) from point $(t, x(t))$ to point $(t+1, x(t+1))$ for a fixed material element. The deformation tensor defined on the material in velocity space and spatial space is:

$$f^i(\eta, x(t))\vec{g}_i(x(t)) \rightarrow f^i(\eta, x(t+1)) \cdot \vec{g}_i(x(t+1)) \tag{27}$$

The instant path deformation tensor $\widetilde{F}^i_j(t, x(t))$ is defined as:

$$\widetilde{F}^i_j = F^i_l F^l_j \tag{28}$$

It is clear that this is the deformation tensor on material motion path. For a fixed material element, taking the instant velocity field at point $(t, x(t))$ as $f_0^i \vec{g}_i^0$ and the instant velocity field at point $(t+1, x(t+1))$ as $f^i \vec{g}_i$, one has the formulation:

$$f^i \vec{g}_i = (F^i_l f_0^l)(F^k_j \vec{g}_k^0) = (F^i_l F^k_i f_0^l) \vec{g}_k^0 \tag{29}$$

Hence, for a fixed material element, when the Eular coordinator system is taken as the measuring system, the velocity component along the material motion path can be expressed as:

$$V^i = F^i_l F^l_j V_0^j = \widetilde{F}^i_j V_0^j \tag{30}$$

In Newton velocity space, the motion path of a material element is defined as a deformation with time parameters. This picture is equally applicable for material element in continuum or the mass point as exterior material moving in continuum. Hence, this deformation tensor is in general sense.

### 3.4 Kinetic Energy and Deformation Energy

For any instant, the kinetic energy is defined as:

$$E(x(t)) = \frac{\rho_0}{2} g_{ij} V^i V^j \tag{31}$$

For isotropic simple continuum, letting its Lame constants are $(\lambda, \mu)$, the stress tensor is defined as:

$$\sigma^i_j = \lambda(F^l_l - 3)\delta^i_j + 2\mu(F^i_j - \delta^i_j) \tag{32}$$

As the stress is defined by the difference between two configurations, the continuum intrinsic physical feature can be defined as:

$$C^{ik}_{jl} = \lambda \delta^i_j \delta^k_l + 2\mu \delta^i_l \delta^k_j \tag{33}$$

The deformation energy of continuum is defined as:

$$U = \sigma^i_j (F^j_i - \delta^j_i) = C^{ik}_{jl}(F^l_k - \delta^l_k)(F^j_i - \delta^j_i) \tag{34}$$

Defining the symmetry parts and anti-symmetry part of deformation tensor $F^i_j$ as:

$$\varepsilon_{ij} = \frac{1}{2}[(F^i_j - \delta^i_j) + (F^j_i - \delta^j_i)] \tag{35-1}$$

$$\omega_{ij} = \frac{1}{2}(F^i_j - F^j_i) \tag{35-2}$$

The deformation energy is in the form:

$$U = \lambda(\varepsilon_{ll})^2 + 2\mu \cdot \varepsilon_{ij}\varepsilon_{ij} - 2\mu \cdot \omega_{ij}\omega_{ij} \tag{36}$$

It shows that the anti-symmetrical part of configuration deformation gives out negative energy. To explain the intrinsic physical meaning of above formulation, consider the mass point motion in constant path velocity.

*3.5 Mass Point Motion with Constant Path Velocity*

The constant path velocity means the kinetic energy is a constant along motion path. In physics tradition, this path is named as geodesic path. In Newton velocity space, the geometrical condition is that:

$$g_{ij} = F^l_i F^k_j g^0_{lk} = F^l_i F^l_j = \delta_{ij} \tag{37}$$

Hence, the continuum deformation is an orthogonal local rotation tensor. For simplicity, without losing generality, let the rotation axe direction is along the $x^3$ path direction (defined as $x^3$ direction). Then, the continuum deformation tensor is simplified as:

$$R^i_j = \begin{vmatrix} \cos\Theta & \sin\Theta & 0 \\ -\sin\Theta & \cos\Theta & 0 \\ 0 & 0 & 1 \end{vmatrix} \tag{38}$$

Using above results, the deformation energy is obtained as:

$$U = 4(\lambda + \mu)(1 - \cos\Theta)^2 - 4\mu \cdot (\sin\Theta)^2 \tag{39}$$

For $\Theta \neq 0$, the deformation energy is existing in the motion path on local rotating plane.

In Newton mechanics, the deformation energy is appointed as zero. What are the physical results by this condition? If $U = 0$, one has:

$$(\lambda + \mu)(1 - \cos\Theta) - \mu \cdot (1 + \cos\Theta) = 0 \tag{40}$$

Its solution is:

$$\cos\Theta_0 = \frac{\mu}{\lambda + 2\mu} \tag{41}$$

Once the continuum features are given, the local rotation angle of motion path is determined. On this motion path, the curvature radium ($R_0 = \frac{1}{\Theta_0}$) of motion path is independent the kinetic energy (which is constant). So, all paths on a sphere with radium $R_0 = \frac{1}{\Theta_0}$ are solutions. The surface of sphere ($R_0 = \frac{1}{\Theta_0}$) defines a geodesic surface.

In fact, zero deformation energy defines the geodesic surface in geodesy. For more general case, the $\lambda = \lambda(x)$ and $\mu = \mu(x)$ are spatial position dependent. In this case, the geodesic surface can be viewed as the surface obtained by sphere surface deformation.

Physically speaking, the geodesic surface (zero deformation energy) divides the spatial continuum into two regions: 1) exterior space and interior space. They have different spatial continuum deformation energy sign.

Reasoning from this point of view, there are two limit geodesic surfaces.

*3.5.1 Black Hole Geodesic Surface*

For gravity field, the black hole is defined as any material will be trapped into it. From motion path view-point, it means the local rotation angle is near $\frac{\pi}{2}$. So, the geometrical field definition of black hole in spatial continuum is:

$$\cos \Theta_0 = \frac{\mu}{\lambda + 2\mu} \to \infty, \quad \Theta_0 \to \frac{\pi}{2} \tag{42}$$

The only possibility is that:

$$\lambda + 2\mu \to 0 \tag{43}$$

In continuum mechanics sense, this means the spatial continuum has no elasticity (not deformable). Based on present knowledge, the $\mu$ is equivalent with gravity mass, then a natural guise about the parameter $\lambda(\to -\frac{m}{2}), (\mu = m)$ is that "it is anti-mass".

In particle physics, when two particles (with limit mass, moving in anti-directions) are combined into one new particle, the effects of $\lambda$ (anti-matter) must be taken into consideration. Personally, I believe that: the high energy physical experiments will expose the final feature of parameter $\lambda$ in electrical field and gravity field.

*3.5.2 Infinite Cosmic Boundary Surface*

As another limit case, when the motion path is an absolute straight line (which is pure Newton spatial continuum), the local rotation angle tends to zero. It defines an infinite cosmic boundary surface. Physically, the condition is that:

$$\cos \Theta_0 = \frac{\mu}{\lambda + 2\mu} \to 0, \quad \Theta_0 \to 0 \tag{44}$$

The two possible solutions are that:

1) Mass-less field, features as:

$$\lambda \neq 0, \quad \mu \to 0 \tag{45}$$

The electromagnetic field meets this condition. So, it is understandable that: any observation of radiating electromagnetic field will lead to conclude that the cosmic has infinite boundary or is expanding state (as it is in Big-bang theory).

2) Anti-matter spatial continuum. As the $\lambda$ is viewed as anti-matter based on current conventional physical theory, another solution is the anti-matter is filled in the cosmic space. It is

featured as:
$$\lambda \to \infty, \text{ for limit } \mu \tag{46}$$

In this case, the spatial continuum has infinite elasticity. So, finite spatial continuum will be compressed into a point-like black-hole is an acceptable conclusion. Also, the finite continuum will be produced by point-like black-hole is an acceptable conclusion.

In fact, the results mentioned above are well-supported by cosmic physical observations.

*3.5.3 Lie algebra as a first order Approximation*

On the geodesic path, the motion equation in Newton velocity form is:

$$\frac{dx^1}{dt} = f_0^1 \cdot \cos\Theta - f_0^2 \cdot \sin\Theta \tag{47-1}$$

$$\frac{dx^2}{dt} = f_0^1 \cdot \sin\Theta + f_0^2 \cos\Theta \tag{47-2}$$

$$\frac{dx^3}{dt} = f_0^3 \tag{47-3}$$

Where, $f_0^i$ are constants. This gives out a close path with a time cycle $T = T(\Theta)$. This is the case in gravity field or in static electrical field. The potential field stress acting on the motion path is defined as:

$$\sigma_j^i = -2\lambda(1-\cos\Theta)\delta_j^i + 2\mu \cdot \begin{vmatrix} \cos\Theta - 1 & \sin\Theta & 0 \\ -\sin\Theta & \cos\Theta - 1 & 0 \\ 0 & 0 & 0 \end{vmatrix} \tag{48}$$

When the $\Theta$ is very small, it is simplified as:

$$\sigma_j^i \approx 2\mu \cdot \begin{vmatrix} 0 & \Theta & 0 \\ -\Theta & 0 & 0 \\ 0 & 0 & 0 \end{vmatrix} = 2\mu \cdot \Theta \cdot \begin{vmatrix} 0 & 1 & 0 \\ -1 & 0 & 0 \\ 0 & 0 & 0 \end{vmatrix} \tag{49}$$

The continuum deformation tensor is approximated as:

$$R_j^i \approx \begin{vmatrix} 1 & \Theta & 0 \\ -\Theta & 1 & 0 \\ 0 & 0 & 1 \end{vmatrix} = \begin{vmatrix} 1 & 0 & 0 \\ 0 & 1 & 0 \\ 0 & 0 & 1 \end{vmatrix} + \Theta \cdot \begin{vmatrix} 0 & 1 & 0 \\ -1 & 0 & 0 \\ 0 & 0 & 0 \end{vmatrix} \tag{50}$$

Comparing with Equation (19), one has:

$$\frac{\partial f^i}{\partial x^j} \approx \Theta \cdot \begin{vmatrix} 0 & 1 & 0 \\ -1 & 0 & 0 \\ 0 & 0 & 0 \end{vmatrix} \tag{51}$$

In Hamilton system, it is the structure matrix of Poisson brackets. Hence, the Lie algebra is a first order approximation.

As the local rotation angle of motion path is the invert of curvature radium $R$ of path, the stress field of Equation (49) is the same as the potential field in form:

$$\sigma_j^i \approx \frac{2\mu}{R} \cdot \begin{vmatrix} 0 & 1 & 0 \\ -1 & 0 & 0 \\ 0 & 0 & 0 \end{vmatrix} \tag{52}$$

Hence, it corresponds to conventional potential field: $\frac{2\mu}{R}$. Then, the viscosity parameter of spatial continuum is cleared.

Based on above understanding, when the first order approximation is taken, the formulation in this paper returns to the Lie algebra description. As the Lie algebra description is conventional content for physicists, we will stop at here without further discussion.

### *3.6 Mass Point Motion in Special Relativity*

In special relativity, the kinetic energy of mass point motion is given as:

$$E = \frac{1}{2}\frac{\rho_0}{\gamma} g_{ij} V^i V^j \tag{53}$$

Where, the $\gamma = \sqrt{1-(\frac{V}{c})^2}$, ($(V)^2 = g_{ij} V^i V^j$), is Lorentz parameter.

If the kinetic energy is conservative, one must has:

$$\frac{g_{ij} V^i V^j}{\sqrt{1-\frac{g_{ij}V^i V^j}{c^2}}} = V_0^i V_0^i = (V_0)^2 = const \tag{54}$$

It is clear that the local rotation meets this equation.

This equation gives out a simple algebra equation about $(V)^2$ as:

$$(V)^4 + \frac{(V_0)^4}{c^2} \cdot (V)^2 - (V_0)^4 = 0 \tag{55}$$

Its roots are expressed as:

$$(V)^2 = \frac{(V_0)^2}{2} \cdot [\frac{(V_0)^2}{c^2} \pm \sqrt{\frac{(V_0)^4}{c^4}+4}] \tag{56}$$

Note that, as $\frac{(V_0)^2}{c^2} < \sqrt{\frac{(V_0)^4}{c^4}+4}$, the negative kinetic energy is a solution. Then, the negative kinetic energy should be understood as the energy relative to the special-time continuum energy definition. Hence, it is a relative quantity. In the following discussion, the $(V_0)^2 > 0$ is supposed.

### *3.6.1 Motion Path Bifurcation*

The positive solution gives out the real path velocity as:

$$V_{real} = \pm V_0 \cdot \sqrt{\frac{1}{2} \cdot [\frac{(V_0)^2}{c^2} + \sqrt{\frac{(V_0)^4}{c^4}+4}]} \tag{57}$$

Its kinetic energy is positive.

The negative solution gives out the pure imaginary velocity solution as:

$$V_{imag} = \pm\sqrt{-1} \cdot V_0 \cdot \sqrt{\frac{1}{2} \cdot [\sqrt{\frac{(V_0)^4}{c^4}+4} - \frac{(V_0)^2}{c^2}]} \tag{58}$$

Its kinetic energy is negative. The negative mass concept may be used to explain it, as it is in conventional physics sense. However, in this research, the negative mass concept is defined by black hole concept (3.5.1 Black Hole Geodesic Surface).

It shows that: in Newton velocity phase space, for any initial state (a sphere surface path), the mass point will jump to one of two states (a sphere surface path): 1) jump to an internal sphere surface path (real path velocity); 2) jump to an external surface path (imaginary path velocity). This is the basic physical picture for atom and molecular physics (the motion path jumping). By Equation (56), the difference of path velocity is $\Delta(V)^2 = (V_0)^2 \cdot \sqrt{\frac{(V_0)^4}{c^4} + 4}$, in low speed motion ($V_0 \ll c$), this quantity is very small. Hence, only for high speed motion (in nuclear physics) the negative kinetic energy solutions can be obtained with significance.

*3.6.2 Spatial Continuum Deformation Energy Bifurcation*

Without losing generality, for spatial continuum deformation:

$$\frac{\partial V}{\partial x^3} = \frac{1}{\sqrt{1-\left(\frac{V}{c}\right)^2}} - 1, \quad \frac{\partial V}{\partial x^1} = 0, \quad \frac{\partial V}{\partial x^2} = 0 \tag{59}$$

The non zero stress components of spatial continuum are:

$$\sigma_1^1 = \sigma_2^2 = \lambda \cdot (\frac{1}{\sqrt{1-(\frac{V}{c})^2}} - 1), \quad \sigma_3^3 = (\lambda + 2\mu) \cdot (\frac{1}{\sqrt{1-(\frac{V}{c})^2}} - 1) \tag{60}$$

The real path velocity solution means spatial continuum expansion. The imaginary path velocity solution means spatial continuum contraction. So, the quantum bifurcation of motion path is caused by the stress in the spatial continuum (or simply says, caused by potential hole).

For pure charge potential field, $\lambda = 0$, the motion force is supplied by the principle stress component along motion path direction: $\sigma_3^3 = 2\mu \cdot (\frac{1}{\sqrt{1-(\frac{V}{c})^2}} - 1)$. It gives out a natural conclusion: any motion force round the motion path may cause quantum bifurcation.

For mass point motion in special relativity theory, the spatial continuum deformation energy is:

$$U = (\lambda + 2\mu) \cdot (\frac{1}{\sqrt{1-(\frac{V}{c})^2}} - 1)^2 \tag{61}$$

It is always positive.

By Hamilton quantity definition of total energy, if the kinetic energy is negative (imaginary path velocity), the zero total energy can be achieved by local positive potential field. For positive kinetic energy (real path velocity), the zero total energy can be achieved by negative potential field. This complement feature of kinetic energy and potential energy is adopted by the Hamilton dynamic system formulation.

For real and imaginary path velocity, the spatial continuum deformation energy has different values. A potential well can be defined as:

$$\Delta U_{well} = \pm(\lambda + 2\mu) \cdot [(\frac{1}{\sqrt{1-(\frac{V_{real}}{c})^2}} - 1)^2 - (\frac{1}{\sqrt{1-(\frac{V_{imag}}{c})^2}} - 1)^2] \tag{62}$$

For mass point motion path jumping among two paths, it defines the space potential energy variation between two motion paths. In quantum mechanics theory of atoms, this energy is radiated out (or absorbed) by from the mass point during its motion path jumping. Generally speaking, for path jumping, the kinetic energy conservation is achieved through the spatial continuum deformation energy variation. Or, as it is exposed in quantum mechanics, the mass point path jumping is produced by potential well in spatial domain.

*3.6.3 Quantum Motion Path*

Further more, the current gauge tensor of spatial continuum is:

$$g_{11} = g_{22} = 1, \quad g_{33} = \frac{1}{1-\left(\frac{V}{c}\right)^2}, \text{ others are zero} \tag{63}$$

For imaginary path velocity, the gauge component along the path is compressed (shortened, $g_{33} < 1$). When the absolute value of imaginary path velocity tends to light velocity $V_{imag} \to \sqrt{-1} \cdot c$, the path direction gauge tends to $\frac{1}{2}$. ($g_{33} = \frac{1}{1-\left(\frac{V_{imag}}{c}\right)^2} \to \frac{1}{2}$) Therefore, the high speed motion of mass point is in quantum motion path (quantum wave picture). Hereafter, the imaginary path velocity is named as quantum motion path.

For real path velocity, the gauge component along the path is elongated (expanded, $g_{33} > 1$). When the path velocity tends to light velocity $V_{real} \to c$, the path spatial gauge of real velocity tends to infinite large. ($g_{33} = \frac{1}{1-\left(\frac{V_{real}}{c}\right)^2} \to \infty$).

*3.6.4 P. Dirac Formulation of Quantum Mechanics*

The general formulation of quantum mechanics based on special relativity is constructed by great scientist P. Dirac (he is the first scientist who introduce negative mass concept. The negative mass concept is equivalent to negative kinetic energy concept).

As the imaginary moment can be defined by the pure imaginary velocity solution, hence, we obtain the P. Dirac Equation:

$$(E)^2 = (E_0)^2 + P^2 c^2 = (E_0 + \sqrt{-1} \cdot Pc)(E_0 - \sqrt{-1} \cdot Pc) \tag{64}$$

Hence, based on special relativity theory, the kinetic energy will allow the imaginary velocity or the imaginary moment. This forms the base for quantum mechanics.

In fact, P. Dirac (based on the Equation (59)) establishes the operator theory of quantum mechanics. Its algebra structure is also similar with Lie algebra. The details have been the contents of standard textbook in quantum mechanics.

*3.7 Additional Discussion*

To end this section, let us re-consider the spatial continuum deformation energy:

$$U = \lambda(\varepsilon_{ll})^2 + 2\mu \cdot \varepsilon_{ij}\varepsilon_{ij} - 2\mu \cdot \omega_{ij}\omega_{ij} \tag{65}$$

For pure negative spatial continuum deformation energy, the Poisson brackets are in the ideal form:

$$U_\omega = -2\mu \cdot \omega_{ij}\omega_{ij} \tag{66}$$

The Lie algebra formulation gives out an exact formulation. The infinitesimal rotation field can be described by $\omega_{ij}$ field. Its general form is Poisson bracket.

When the symmetry parts of spatial continuum deformation play the main role, if one absolutely reasoning with Lie algebra formulation, he must require that:

$$U_\varepsilon = \lambda(\varepsilon_{ll})^2 + 2\mu \cdot \varepsilon_{ij}\varepsilon_{ij} = 0 \tag{67}$$

Its natural conclusion is that: $\lambda + 2\mu = 0$. Then, the anti-matter concept must be introduced by instinct reasoning.

Is that the facts happening in modern physics theory?

Summing the related results in this section, the Hamilton quantity for mass point motion along its path is:

$$H = E + U = \frac{\rho}{2} g_{ij} V^i V^j + \lambda(\varepsilon_{ll})^2 + 2\mu \cdot \varepsilon_{ij}\varepsilon_{ij} - 2\mu \cdot \omega_{ij}\omega_{ij} \tag{68}$$

For constant Hamilton quantity system, that is the energy conservation system.

1). Atom System. If $U_\varepsilon = 0$, the infinitesimal rotation field is equivalent with self-spin. Then the Atom system can be defined as:

$$H = \frac{\rho}{2} g_{ij} V^i V^j - 2\mu \cdot \omega_{ij}\omega_{ij} = const \tag{69}$$

This describes the atom system. The kinetic energy will not be radiated out from the system. The atom is a stable dynamic system. Hence, the Poisson brackets are good tools.

For $H = 0$, if the kinetic energy is positive, the $\mu$ must be positive. If the kinetic energy is negative, the $\mu$ must be negative. Hence, there is an intrinsic symmetry about mater features. This role is taken up by positive electronic charge and negative electronic charge in atom dynamic system.

If one wants to extend this model to cosmic system, the anti-matter is the looking for item.

2). Vacuum. If the vacuum is defined by the Hamilton quantity, the natural definition of vacuum definition is zero Hamilton quantity, that is:

$$H = E + U = \frac{\rho}{2} g_{ij} V^i V^j + \lambda(\varepsilon_{ll})^2 + 2\mu \cdot \varepsilon_{ij}\varepsilon_{ij} - 2\mu \cdot \omega_{ij}\omega_{ij} = 0 \tag{70}$$

For an observer in commoving system, the kinetic energy item can be eliminated. In this case, the spatial continuum is in a dynamic deforming process.

$$\lambda(\varepsilon_{ll})^2 + 2\mu \cdot \varepsilon_{ij}\varepsilon_{ij} - 2\mu \cdot \omega_{ij}\omega_{ij} = 0 \tag{71}$$

For $\mu > 0$, one must have $\lambda < 0$; For $\mu < 0$, one must have $\lambda > 0$. The intrinsic symmetry of matter is maintained.

In fact, all above topics have been extensively studied in modern physics. Here, the related results will not be repeated.

Physically, for the motion of a matter element, the Hamilton quantity shows that the total energy is the addition of kinetic energy and potential field energy where the matter is in motion. As the zero reference point of potential energy is arbitral, the zero point of Hamilton quantity is arbitral also. Only based on this understanding, the negative kinetic energy can be explained.

To remove the arbitrary of zero reference point of potential energy, the absolute energy concept can be introduced. The absolute energy reference function $h(T)$ is defined by the absolute temperature $T$ in modern physics. As the absolute temperature is positive, the general Hamilton quantity should be defined as:

$$\tilde{H} = E + U + h(T) = \frac{\rho}{2} g_{ij} V^i V^j + \lambda(\varepsilon_{ll})^2 + 2\mu \cdot \varepsilon_{ij}\varepsilon_{ij} - 2\mu \cdot \omega_{ij}\omega_{ij} + h(T) \qquad (72)$$

Where, the $h(T)$ defines the absolute background of time-spatial continuum. Therefore, the absolute zero Hamilton quantity can exist (vacuum in physical sense). Based on physical reality consideration, the absolute Hamilton quantity must be positive.

In classical thermodynamics, the absolute temperature $T$ is defined by equation:

$$h(T) = \frac{\rho}{2} g_{ij} V^i V^j \qquad (73)$$

Where, the velocity $V^i$ is Newton velocity.

It is clear that for zero Hamilton point, the definition of absolute temperature should take the spatial continuum deformation into consideration.

Above discussion is made to show that: the negative kinetic energy is physical reality predicted by the Hamilton dynamic system.

In conventional selection, the negative kinetic energy should be understood as radiation. By this sense, the material element motion in potential field has two possible motions: 1) positive kinetic energy (classical Newton motion); 2) negative kinetic energy, the radiating motion.

## 4. General Motion Equations

In previous sections, the geometrical field description developed in this research is that: the basic physical realities are explained in an united formulation. The abstract intrinsic physical features of spatial continuum are described by two basic physical parameters. The most important thing is that: in the geometrical field formulation, the quantum mechanics formulation is obtained as a natural result of special relativity theory without requiring external assumption. Further, the geometrical motion in spatial path and the kinetic motion in phase path are formulated by the same deformation tensor.

As the effectiveness of geometrical field description is explained in above sections, in this section, general motion equations will be studied.

### *4.1 Stress Tensor and Acceleration Force*

For a general dynamic system in classical form defined in inertial coordinator system $(x^1, x^2, x^3; \vec{g}_1^0, \vec{g}_2^0, \vec{g}_3^0)$:

$$\frac{dx^i}{dt} = f^i(\eta, x) \qquad (74)$$

The instant deformation of continuum (where the motion is happening) is defined as:

$$F_j^i(\eta, x) = \delta_j^i + \frac{\partial f^i(\eta, x)}{\partial x^j} \qquad (75)$$

It defines the local manifold configuration deformation. For simple isotropic continuum ($\lambda, \mu$), the stress in the local continuum is defined as:

$$\sigma_j^i = \lambda(F_l^l - 3)\delta_j^i + 2\mu(F_j^i - \delta_j^i) \qquad (76)$$

It is a surface force defined on the surface formed by the material mass element traces. (Here, the point mass is defined by unit volume element. Hence, its trace is a tube rather than a mathematic line).

If Hamilton quantity is used, the stress is: $\sigma_j^i = \lambda(\frac{\partial^2 H}{\partial p_i \partial x^l} - 3)\delta_j^i + 2\mu(\frac{\partial^2 H}{\partial p_i \partial x^j} - \delta_j^i)$.

On the other hand, when the configuration is deformed as:

$$\vec{g}_i(x(t+1)) = F_i^{\,j}(\eta, x(t)) \cdot \vec{g}_j(x(t)) \tag{77}$$

The path velocity (absolute velocity along the trace path) is transformed as:

$$V^i(t+1, x(t+1)) = F_l^{\,i}(\eta, x(t)) \cdot F_j^{\,l}(\eta, x(t)) \cdot V^{\,j}(t, x(t)) = \tilde{F}_j^{\,i}(\eta, x(t)) \cdot V^{\,j}(t, x(t)) \tag{78}$$

Where, $\tilde{F}_j^{\,i}(\eta, x(t)) = F_l^{\,i}(\eta, x(t)) \cdot F_j^{\,l}(\eta, x(t))$.

It gives out the acceleration along its trace path as:

$$\Delta a^i(t, x) = [\tilde{F}_j^{\,i}(\eta, x) - \delta_j^i] \cdot V^{\,j}(t, x(t)) \tag{79}$$

Hence, the acceleration force (body force) acting on the material element is:

$$F^i(t, x(t)) = \rho[\tilde{F}_j^{\,i}(\eta, x(t)) - \delta_j^i] \cdot V^{\,j}(t, x(t)) \tag{80}$$

Where, the $\rho$ is the mass of material element under discussion.

By definition equations:

$$\vec{V}(t, x(t)) = \frac{dx^i(t)}{dt} \cdot \vec{g}_i(x(t)) = f^{\,i}(\eta, x(t)) \cdot \vec{g}_i(x(t)) \tag{81}$$

Along the motion path, the acceleration force (body force) acting on the material element is expressed as:

$$F^i(t, x(t)) = \rho \cdot [\tilde{F}_j^{\,i}(\eta, x(t)) - \delta_j^i] \cdot f^{\,j}(\eta, x(t)) \tag{82}$$

Similar with the reasoning in continuum mechanics, the force balance equations [12] along the trace path are:

Linear Momentum Conservation Equation:

$$\sigma_j^i(x)\big|_j = \rho \cdot [\tilde{F}_j^{\,i}(\eta, x) - \delta_j^i] \cdot f^{\,j}(\eta, x) \tag{83}$$

Angular Momentum Conservation Equation:

$$\sigma_j^i(x)\big|_i = \rho \cdot [\tilde{F}_l^{\,k}(\eta, x) - \delta_l^k] \cdot f^{\,l}(\eta, x) \cdot g_{ki}(x) \cdot F_j^{\,i}(\eta, x) \tag{84}$$

Consistency Equation:

$$e_{ijk} F_l^{\,j} \sigma_k^{\,l} = 0 \tag{85}$$

The Equation (85) is met naturally by the stress definition Equation (76). For other stress form, this equation is must be met one.

Once the stress definition meets Equation (85), for the other two sets of equations, the wanted quantity is $f^{\,i}(\eta, x)$ (or $\frac{dx^i}{dt}$) functions and $g_{ij}(x)$. (Please note that the direct time-dependence is removed from above equations. This fact is used to construct general theory in phase space in modern physics. )

The above equations show that the initial features of continuum will take a role in the final solution. Hence, in this sense, the potential field determines a specific spatial feature. This feature is implied in the general form of motion equation (Equation 74). In mathematic sense, the dynamic system is over-determined. This over determined feature makes the physical parameters are limited by special values. That is they are discrete solutions.

On physical sense, for given functions $f^i(\eta, x)$ and $g_{ij}^0$ space, there are three physical parameters: $\lambda, \mu, \rho$. The equations determine their allowable values. In inverse is true.

So, the Hamilton dynamic system will gives out the form solutions:

$$(\lambda_\alpha, \mu_\alpha, \rho_\alpha), \alpha = 1,2,...; \quad f^i(\eta_\alpha, x), \alpha = 1,2,....; i = 1,2,3 \tag{86}$$

The conclusion is that: the Hamilton dynamic system is a quantum system.

When only taking Equation (83), the system is degenerated to conventional Newton system, the solution is not quantum.

When only taking Equation (84), the system is degenerated to conventional Eular system, the solution is not quantum.

When only taking Equation (85), symmetry becomes the essential features. This is very hot in modern physics research.

When only omitting one set of the equation, part solution of quantum can be obtained. This is the fact happens in modern physics.

Although at present stage, the author cannot supply enough evidence to support above results, a possible way is formulated for connecting the classical physics with quantum physics in this research.

*4.2 Constant Absolute Velocity*

Without losing generality, taking the initial coordinator system as standard rectangular system $g_{ij}^0 = \delta_{ij}$ (means observing with standard measurement system), for the motion with constant absolute velocity $g_{ij} V^i V^j = Const$, the condition is that: $g_{ij} = F_i^l F_j^k g_{lk}^0 = \delta_{ij}$. In general case, one has:

$$F_j^i = S_j^i + R_j^i(\Theta) \tag{87-1}$$

Where, the $S_j^i$ is a symmetrical tensor, the $R_j^i(\Theta)$ is an unit orthogonal rotation tensor with rotation angular $\Theta$. In standard rectangular coordinator system, they are defined as:

$$S_j^i = \frac{1}{2}(\frac{\partial f^i}{\partial x^j} + \frac{\partial f^j}{\partial x^i}) - (1 - \cos\Theta) L_l^i L_j^l \tag{87-2}$$

$$R_j^i = \delta_j^i + \sin\Theta \cdot L_j^i + (1 - \cos\Theta) L_l^i L_j^l \tag{87-3}$$

$$L_j^i = \frac{1}{2\sin\Theta}(\frac{\partial f^i}{\partial x^j} - \frac{\partial f^j}{\partial x^i}) \tag{87-4}$$

$$(\sin \Theta)^2 = \frac{1}{4}[(\frac{\partial f^1}{\partial x^2} - \frac{\partial f^2}{\partial x^1})^2 + (\frac{\partial f^2}{\partial x^3} - \frac{\partial f^3}{\partial x^2})^2 + (\frac{\partial f^3}{\partial x^1} - \frac{\partial f^1}{\partial x^3})^2] \quad (87\text{-}5)$$

Where, the anti-symmetrical tensor $L^i_j$ defines the local rotating axe direction. Please note that: by conventional Hamilton quantity, $L^i_j = \frac{1}{2\sin\Theta}(\frac{\partial^2 H}{\partial p_i \partial x^j} - \frac{\partial^2 H}{\partial p_j \partial x^i})$.

So, the constant absolute velocity means that: $S^i_j \equiv 0$, $F^i_j = R^i_j(\Theta)$.

Then, the path velocity transformation is:

$$V^i(t+1, x(t+1)) = R^i_l(\Theta) \cdot R^l_j(\Theta) \cdot V^j(t, x(t)) = R^i_j(2\Theta) \cdot V^j(t, x(t)) \quad (88)$$

The related motion equation becomes:

$$\sigma^i_j(x)\big|_j = \rho \cdot [R^i_j(2\Theta) - \delta^i_j] \cdot f^j(\eta, x) \quad (89)$$

$$\sigma^i_j(x)\big|_i = \rho \cdot [R^i_l(2\Theta) - \delta^i_l] \cdot f^l(\eta, x) \cdot R^i_j(\Theta) \quad (90)$$

Putting the Equation (89) into Equation (90), one has:

$$\sigma^i_j(x)\big|_i = \sigma^i_l(x)\big|_l \cdot R^i_j(\Theta) \quad (91)$$

Its acceleration force form is:

$$F_j(x) = F^i(x) \cdot R^i_j(\Theta) \quad (92)$$

(This is distinguished from Riemann geometry, as the Riemann geometrical equations gives out the equation: $F_j = F^i g_{ij} = F^i R^l_i(\Theta) R^l_j(\Theta) = F^j$. Hence, the index raising or index lowing rules are not applicable for the acceleration force components. This is acceptable as the Riemann geometry is suitable for physical quantities in fixed configuration without deformation.)

For infinitesimal rotation angular, the Equations (89) and (90) can be simplified as one equation:

$$\sigma^i_j(x)\big|_j = \rho \cdot [R^i_j(2\Theta) - \delta^i_j] \cdot f^j(\eta, x), \quad \sigma^i_j = \sigma^j_i \quad (93)$$

The right side of the equation is approximated a Poison brackets (or Lie algebra) as it is explained in previous sections.

*4.3 Variation Absolute Velocity*

Along the material element motion path, the variation absolute velocity can be expressed by a conformal transformation:

$$g_{ij} = \left(\frac{1}{\cos\theta}\right)^2 \delta_{ij} \quad (94)$$

The corresponding instant deformation tensor of continuum is:

$$F^i_j = \frac{1}{\cos\theta} \tilde{R}^i_j(\theta), \quad \tilde{S}^i_j \equiv 0 \quad (95\text{-}1)$$

Where, In standard rectangular coordinator system, the related quantities are defined as:

$$\tilde{S}^i_j = \frac{1}{2}(\frac{\partial f^i}{\partial x^j} + \frac{\partial f^j}{\partial x^i}) - (\frac{1}{\cos\theta} - 1)(\tilde{L}^i_l \tilde{L}^l_j + \delta^i_j) \tag{95-2}$$

$$\tilde{R}^i_j = \delta^i_j + \sin\theta \cdot \tilde{L}^i_j + (1-\cos\theta)\tilde{L}^i_l \tilde{L}^l_j \tag{95-3}$$

$$\tilde{L}^i_j = \frac{\cos\theta}{2\sin\theta}(\frac{\partial f^i}{\partial x^j} - \frac{\partial f^j}{\partial x^i}) \tag{95-4}$$

$$(\frac{1}{\cos\theta})^2 = 1 + \frac{1}{4}[(\frac{\partial f^1}{\partial x^2} - \frac{\partial f^2}{\partial x^1})^2 + (\frac{\partial f^2}{\partial x^3} - \frac{\partial f^3}{\partial x^2})^2 + (\frac{\partial f^3}{\partial x^1} - \frac{\partial f^1}{\partial x^3})^2] \tag{95-5}$$

Where, the anti-symmetrical tensor $\tilde{L}^i_j$ defines the local rotating axe direction.

The path velocity transformation is:

$$V^i(t+1, x(t+1)) = \left(\frac{1}{\cos\theta}\right)^2 \tilde{R}^i_j(2\theta) \cdot V^j(t, x(t)) \tag{96}$$

The related motion equation becomes:

$$\sigma^i_j(x)\big|_j = \rho \cdot [\left(\frac{1}{\cos\theta}\right)^2 \tilde{R}^i_j(2\theta) - \delta^i_j] \cdot f^j(\eta, x) \tag{97}$$

$$\sigma^i_j(x)\big|_i = \rho \cdot [\left(\frac{1}{\cos\theta}\right)^2 \tilde{R}^i_l(2\theta) - \delta^i_l] \cdot f^l(\eta, x) \cdot \frac{1}{\cos\theta} \cdot \tilde{R}^i_j(\theta) \tag{98}$$

They give out the relation equation:

$$\sigma^i_j(x)\big|_i = \sigma^i_l(x)\big|_l \cdot \frac{1}{\cos\theta} \tilde{R}^i_j(\theta) \tag{99}$$

Or, in its acceleration force form as:

$$F_j(x) = F^i(x) \cdot \frac{1}{\cos\theta} \tilde{R}^i_j(\theta) \tag{100}$$

As a simple result of observation, the path velocity vector and acceleration force vector are not the vectors defined in Riemann geometry.

By Equations (92) and (100), for particle motion in potential fields, the Poisson brackets or Lie algebra is a natural selection to define the velocity field and force field. However, they are first approximation. Hence, the rotation is limited as infinitesimal.

### *4.4 Quantum Solutions*

Summering above two typical motions along paths, for given spatial local rotation angular within unite time, two features are very significant.

1) For constant absolute velocity, referring to its initial configuration, the configuration after unit time is transformed as:

$$\vec{g}_i(x(t+1)) = R^j_i(\Theta) \cdot \vec{g}_j(x(t)) \tag{101}$$

The path velocity component is transformed as:

$$V^i(t+1, x(t+1)) = R^i_j(2\Theta) \cdot V^j(t, x(t)) \tag{102}$$

For enough time duration, the trace is periodic. Since the local rotation angular is defined for unit-time duration, the period of trace in configuration space and the period of phase can be

defined as:

$$T_{trace} = \frac{2\pi}{\Theta}, \quad T_{phas} = \frac{\pi}{\Theta} = 2T_{trace} \quad (103)$$

Using these parameters, the periodic features are expressed as:

$$\vec{g}_i(x(t + \frac{2n\pi}{\Theta})) = \vec{g}_i(x(t)), \quad n = 1, 2, \ldots \quad (104)$$

$$V^i(t + \frac{n\pi}{\Theta}, x(t + \frac{n\pi}{\Theta})) = \cdot V^i(t, x(t)), \quad n = 1, 2, \ldots \quad (105)$$

2) For variation absolute velocity, referring to its initial configuration the configuration after time duration $T$, the configuration is transformed as:

$$\vec{g}_i(x(t+1)) = \left(\frac{1}{\cos\theta}\right) \cdot \tilde{R}_i^j(\theta) \cdot \vec{g}_j(x(t)) \quad (106)$$

The path velocity component is transformed as:

$$V^i(t+1, x(t+1)) = \left(\frac{1}{\cos\theta}\right)^2 \tilde{R}_j^i(2\theta) \cdot V^j(t, x(t)) \quad (107)$$

The period of trace in configuration space and the periodic of phase can be defined as:

$$\tilde{T}_{trace} = \frac{2\pi}{\theta}, \quad \tilde{T}_{phas} = \frac{\pi}{\theta} = 2\tilde{T}_{trace} \quad (108)$$

Using these parameters, after $n$ time period of time duration, the periodic features are expressed as:

$$\vec{g}_i(x(t + \frac{2n\pi}{\theta})) = \left(\frac{1}{\cos\theta}\right)^{\pm\frac{2n\pi}{\theta}} \vec{g}_i(x(t)), \quad n = 1, 2, \ldots \quad (109)$$

$$V^i(t + \frac{n\pi}{\theta}, x(t + \frac{n\pi}{\theta})) = \left(\frac{1}{\cos\theta}\right)^{\pm\frac{n\pi}{\theta}} \cdot V^i(t, x(t)), \quad n = 1, 2, \ldots \quad (110)$$

Where, the positive $n$ corresponds to velocity increase continuously, the negative $n$ corresponds to velocity decrease continuously (it is obtained by reverse time direction). It means that, for enough long time duration, zero space and infinite space are two extremes. The big-bang and expansion cosmic model is an example for such a kind of solution for positive $n$. The black-hole is another example for negative $n$.

Hence, in wave motion terms, the trace angular frequencies are $\Theta$ and $\theta$ respectively, while the phase angular frequencies are $2\Theta$ and $2\theta$ respectively. The double frequency means that for a Hamilton system, the geometrical picture (displacement field description or position description) is different from the physical picture (velocity field description or moment field description). So, in the Hamilton system description, the position variant and moment variant are viewed as independent in formulation. Their intrinsic relation is formulated by Lie algebra formulation.

For the constant absolute velocity, the trace is simply on a sphere surface like an electron motion around a nuclear center. There are no singular points. Hence, the motion is stable.

However, for the variation absolute velocity case, if the unit time is bigger than the trace period (for natural time unit: second, this is the case in quantum mechanics), within a unit time in configuration space, there are two singular points in path velocity space:

$$\frac{1}{\cos\theta} \to \infty, \quad \theta \to \frac{\pi}{2}, \frac{3\pi}{2} \tag{111}$$

If a physical interpretation should be made, the only interpretation is that: the material element is radiated from the spatial continuum. Hence, it should be viewed as a radiating solution.

Hence, defining the highest angular frequency of radiation solution as $\omega_0 = \frac{2\pi}{T_{trace}}$ in configuration space, for variation absolute velocity case, the radiation solutions in configuration space are defined by the following angular frequencies:

$$\omega_0, \; \frac{\omega_0}{3}, \; \frac{\omega_0}{5}, \; \ldots = \frac{\omega_0}{(2n+1)}, \quad n=1,2,\ldots \tag{112}$$

Similarly, in the path velocity space (phase space), defining the lowest angular frequency of radiation solution as $\tilde{\omega}_0 = \frac{\pi}{T_{trace}} = \frac{\omega_0}{2}$, the radiation solutions in phase space are defined by the following angular frequencies:

$$\frac{\omega_0}{2}, \; \frac{1}{3}\cdot\frac{\omega_0}{2}, \; \frac{1}{5}\cdot\frac{\omega_0}{2}, \; \ldots = \frac{1}{(2n+1)}\cdot\frac{\omega_0}{2}, \quad n=1,2,\ldots \tag{113}$$

Based on physical consideration, for low angular frequency, the radiation is degenerated as peak-valley energy alternating processes. The continuous motion picture is recovered.

They show that, in configuration space and in phase space, the observed singular frequencies are different. For the path velocity variation, it means that a path jump and the path velocity jump is not in-phase.

By selecting the period as the unit time, the $\frac{1}{\cos\theta}$ will pass its singular point $\theta = \frac{\pi}{2}$. Now we ask a question: for very small path variation defined by $\delta\theta$ around a singular point in phase space, how the trace kinetic energy is varied?

For a path jumping, if the kinetic energy of a material element is varied from a stable path $\theta_0$ with kinetic energy $E_0$ to a very near path $\theta_0 + \delta\theta$, the kinetic energy variation is limited as $\delta E_{path}$ for path variation. For a dynamic path jumping, taking the period as unit time, viewing Equation (107), the kinetic energy related with path variation should be defined by the limit variation of local rotation angular near a singular point as:

$$\delta E_{path} = E_0 \int_{\frac{\pi}{2}-\delta\theta}^{\frac{\pi}{2}+\delta\theta} (\frac{1}{\cos\theta})^2 d\theta = 2E_0 \tan\frac{\delta\theta}{2} \tag{114}$$

From physical consideration, for very small radiation energy (path perturbation), $\frac{\delta E_{path}}{E_0} \ll 1$, noting that $\delta\theta \propto \delta\omega$, one has:

$$\delta E_{path} \propto \cdot\delta\omega \tag{115}$$

For vacuum continuum with suitable space and time scale as its time-spatial features, its energy $E_0$ should be also fixed. Based on present physics observation, the minimum radiating energy unit is $\hbar$. When a suitable time-space scale is referred, the absolute energy $E$ and the path frequency $\omega$ has the general relationship in statistic sense of the form:

$$E = \hbar \cdot \omega \tag{116}$$

This equation is a basic stone in modern physics. It takes the zero frequency as the zero energy point. By this way, the zero frequency is defined by rigid space-time continuum. Hence, the vacuum has zero point energy.

However, for Hamilton dynamic system, the measurable quantities are the kinetic energy variation (refer Equation (111)). Hence, the average of kinetic energy for all paths is defined as the reference energy (zero energy). Once the average level is taken as the zero point, the negative kinetic energy is naturally introduced. For the negative kinetic energy, a natural way is to introduce imaginary time coordinator. Hence, in quantum mechanics, the time coordinator is imaginary.

Recalling that, at previous section (3.6.1 Motion Path Bifurcation), the negative kinetic energy is obtained by the motion equation of special relativity. Hence, the special relativity theory and quantum mechanics theory are logic consistent. It also shows that, taking the Hamilton dynamic system formulation as the foundation of modern physics is on right way. The only problem is how to make it strong and accurate. This is the target of this research. To clear the broad effectiveness of Hamilton dynamic system, the global solutions are studied bellow.

### 4.5 Global Solutions of Constant Absolute Velocity Motion

For constant absolute velocity motion, the motion path transformation is defined as:

$$F_j^i = R_j^i(\Theta) \tag{113}$$

Without losing generality, selecting the standard rectangular coordinator system, taking the rotation along $x^3$ direction, then the motion path transformation can be expressed as:

$$R_j^i(\Theta) = \begin{vmatrix} \cos\Theta & \sin\Theta & 0 \\ -\sin\Theta & \cos\Theta & 0 \\ 0 & 0 & 1 \end{vmatrix} \tag{114}$$

As the anti-symmetry stress defines the quantum solutions, here, they are omitted. The symmetry stress tensor components in spatial continuum with parameters $\lambda, \mu$ are given as:

$$\sigma_j^i = -2\lambda(1-\cos\Theta)\delta_j^i + 2\mu \cdot \begin{vmatrix} \cos\Theta-1 & 0 & 0 \\ 0 & \cos\Theta-1 & 0 \\ 0 & 0 & 0 \end{vmatrix} \tag{115}$$

On the motion plane ($x^1, x^2$), the Newton force components can be obtained by the as:

$$F^1 = -2(\lambda+\mu)\cdot\sin\Theta\cdot\frac{\partial\Theta}{\partial x^1} \tag{116-1}$$

$$F^2 = -2(\lambda+\mu)\cdot\sin\Theta\cdot\frac{\partial\Theta}{\partial x^2} \tag{116-2}$$

The path velocity field is defined as:

$$\begin{cases} V^1 = \cos 2\Theta \cdot V_0^1 + \sin 2\Theta \cdot V_0^2 \\ V^2 = \cos 2\Theta \cdot V_0^2 - \sin 2\Theta \cdot V_0^1 \end{cases} \tag{117}$$

Where, the $V_0^i$ are constants (initial path velocity). For very small $\Theta$, the motion equations are:

$$-2(\lambda + \mu) \cdot \sin \Theta \cdot \frac{\partial \Theta}{\partial x^1} = \rho \sin 2\Theta \cdot V_0^2 \qquad (118\text{-}1)$$

$$2(\lambda + \mu) \cdot \sin \Theta \cdot \frac{\partial \Theta}{\partial x^2} = \rho \sin 2\Theta \cdot V_0^1 \qquad (118\text{-}2)$$

It can be approximated as:

$$-(\lambda + \mu) \cdot \frac{\partial \Theta}{\partial x^1} = \rho \cdot V_0^2 \qquad (119\text{-}1)$$

$$(\lambda + \mu) \cdot \frac{\partial \Theta}{\partial x^2} = \rho \cdot V_0^1 \qquad (119\text{-}2)$$

For periodic motion path, the conventional form equations are in Furrier form:

$$\frac{\partial \Theta}{\partial x^1} = \sqrt{-1} \cdot k_1 \cdot \Theta, \quad \frac{\partial \Theta}{\partial x^2} = \sqrt{-1} \cdot k_2 \cdot \Theta \qquad (120)$$

Defining the following quantities:

$$k = \sqrt{(k_1)^2 + (k_2)^2}, \quad V_0 = \sqrt{(V_0^1)^2 + (V_0^2)^2} \qquad (121)$$

The equation is rewritten as the momentum $p$ definition equation:

$$p = \rho \cdot V_0 = (\lambda + \mu) \cdot \Theta \cdot k \qquad (122)$$

It shows that: the Newton momentum definition can be expressed by motion path continuum parameters. Therefore, the differences between the classical definition and quantum definition are eliminated as they are the same in physics.

For a given space continuum (vacuum continuum), its time-spatial feature is fixed by parameter $\Theta_{vacu}$, and $\lambda_0, \mu_0$, hence, in vacuum continuum, one has:

$$p = \hbar \cdot k \qquad (123)$$

Where, the Plank physical constant for vacuum continuum is expressed as:

$$\hbar = (\lambda_0 + \mu_0) \cdot \Theta_{vacu} \qquad (124\text{-}1)$$

Or in vacuum Hamilton quantity $H_0(\eta, x, p)$ form:

$$\hbar = \frac{1}{2}(\lambda_0 + \mu_0)\sqrt{(\frac{\partial^2 H_0}{\partial p_1 \partial x^2} - \frac{\partial^2 H_0}{\partial p_2 \partial x^1})^2 + (\frac{\partial^2 H_0}{\partial p_2 \partial x^3} - \frac{\partial^2 H_0}{\partial p_3 \partial x^2})^2 + (\frac{\partial^2 H_0}{\partial p_3 \partial x^1} - \frac{\partial^2 H_0}{\partial p_1 \partial x^3})^2} \qquad (124\text{-}2)$$

Summering above results, the global solution of spatial continuum gives out general equation: $p = \hbar \cdot k$, and the periodic path motion gives out general equation $\varepsilon = \hbar \cdot \omega$.

Hence, the basic equations of quantum mechanics originate from the basic physical features and geometrical features of space-time continuum.

In one words, the Newton mechanics takes the particle motion as its objective picture, while the quantum mechanics takes the motion path variation as its object picture.

### *4.6 Newton Motion Pictures of Mass Potion in Potential Field*

Given a classical dynamic system defined by the velocity vector $\frac{dx^i}{dt} \vec{g}_i$, $i = 1,2,3$ in commoving dragging coordinator system, the conventional motion equation is defined as:

$$\frac{dx^i}{dt} = f^i(\eta, x) \qquad (125)$$

### *4.6.1. Constant Absolute Velocity Motion Picture: Screw Curve*

For constant absolute velocity motion discussed above, in Newton inertial coordinator system ($X,Y,Z$), the mass instant motion direction on periodic path is determined by the following

equation:

$$\vec{g}_X(t) = \vec{g}_X(0) \cdot \cos(\Theta t) + \vec{g}_Y(0) \cdot \sin(\Theta t) \qquad (126\text{-}1)$$

$$\vec{g}_Y(t) = -\vec{g}_X(0) \cdot \sin(\Theta t) + \vec{g}_Y(0) \cdot \cos(\Theta t) \qquad (126\text{-}2)$$

$$\vec{g}_Z(t) = \vec{g}_Z(0) \qquad (126\text{-}3)$$

The path velocity components are observed as:

$$V_X(t) = V_X(0) \cdot \cos(2\Theta t) + V_Y(0) \cdot \sin(2\Theta t) \qquad (127\text{-}1)$$

$$V_Y(t) = -V_X(0) \cdot \sin(2\Theta t) + V_Y(0) \cdot \cos(2\Theta t) \qquad (127\text{-}2)$$

$$V_Z(t) = V_Z(0) \qquad (127\text{-}3)$$

It shows that the original point of Newton inertial coordinator system has a constant absolute velocity $V_X(0), V_Y(0), V_Z(0)$. In fixed position, the path is observed as a screw curve.

Please note that the $\Theta = \Theta[f^i(\Theta, x)]$ and Equation (125) is must be met, so, path velocity cannot be obtained directly from Equation (126) by taking time derivative. The Hamilton system is a physical system, the mathematic derivative rules are determined by it. The physical faith is that: the mathematic rules are determined by physics rules.

The required $\Theta$ defines the potential field features. According to previous definition, for this kind of motion, the $\dfrac{dx^i}{dt} = f^i(\Theta, x)$ in standard rectangular system meets the fowling condition:

$$\frac{\partial f^1}{\partial x^1} = \cos\Theta - 1, \quad \frac{\partial f^2}{\partial x^2} = \cos\Theta - 1, \quad \frac{\partial f^3}{\partial x^3} = 0 \qquad (128\text{-}1)$$

$$\frac{\partial f^1}{\partial x^2} = -\frac{\partial f^2}{\partial x^1} = \sin\Theta, \text{ others are zero} \qquad (128\text{-}2)$$

For constant spatial feature, $\Theta$ is constant. Its solution of Hamilton dynamic system is:

$$\frac{dx^1}{dt} = f^1(\Theta, x(t)) = (\cos\Theta - 1) \cdot x^1(t) + \sin\Theta \cdot x^2(t) + C_1 \qquad (129\text{-}1)$$

$$\frac{dx^2}{dt} = f^2(\Theta, x(t)) = -\sin\Theta \cdot x^1(t) + (\cos\Theta - 1) \cdot x^2(t) + C_2 \qquad (129\text{-}2)$$

$$\frac{dx^3}{dt} = f^3(\Theta, x(t)) = C_3 \qquad (129\text{-}3)$$

Where, the $C_i$ are constants.

For infinitesimal $\Theta$ (weak potential field), as a Newton space approximation, they are simplified as:

$$\frac{dx^1}{dt} = \sin\Theta \cdot x^2(t) + C_1 \qquad (130\text{-}1)$$

$$\frac{dx^2}{dt} = -\sin\Theta \cdot x^1(t) + C_2 \qquad (130\text{-}2)$$

$$\frac{dx^3}{dt} = C_3 \qquad (130\text{-}3)$$

Introducing the curvature radium $R_0 = \dfrac{1}{\Theta}$ as spatial curvature parameter, the Newton initial system is expressed as:

$$\frac{dx^1}{dt} = \frac{x^2(t)}{R_0} + C_1 \qquad (131\text{-}1)$$

$$\frac{dx^2}{dt} = -\frac{x^1(t)}{R_0} + C_2 \tag{131-2}$$

$$\frac{dx^3}{dt} = C_3 \tag{131-3}$$

The ideal Newton inertial system is defined by letting: $R_0 \to \infty$. The initial original point with constant velocity $C_1, C_2, C_3$ is defined.

The space direction variation and absolute velocity in ideal Newton inertial system become:

$$\vec{g}_X(t) = \vec{g}_X(0) \cdot \cos(\frac{t}{R_0}) \tag{126-1}$$

$$\vec{g}_Y(t) = \vec{g}_Y(0) \cdot \cos(\frac{t}{R_0}) \tag{126-2}$$

$$\vec{g}_Z(t) = \vec{g}_Z(0) \tag{126-3}$$

The path velocity components are observed as:

$$V_X(t) = V_X(0) \cdot \cos(\frac{2t}{R_0}) \tag{127-1}$$

$$V_Y(t) = V_Y(0) \cdot \cos(\frac{2t}{R_0}) \tag{127-2}$$

$$V_Z(t) = V_Z(0) \tag{127-3}$$

Comparing with the results in special relativity for free particles, $P_0^2 = P^2(1 - \frac{V_0^2}{c^2})$, for unit time, one obtains that:

$$\cos 2\Theta = \cos\frac{2}{R_0} = \sqrt{\frac{1}{1 - \frac{V_0^2}{c^2}}} = \gamma \tag{128}$$

Hence, the Lorentz parameter $\gamma$. For very long time, the periodic features of space are apparent.

*4.6.2. Variation Absolute Velocity Motion Picture: Spiral Curve*

For variation absolute velocity motion discussed above, in Newton inertial coordinator system ($X, Y, Z$), the mass instant motion direction on periodic path is determined by the following equation:

$$\vec{g}_X(t) = \left(\frac{1}{\cos\theta}\right)^t \cdot [\vec{g}_X(0) \cdot \cos(\theta t) + \vec{g}_Y(0) \cdot \sin(\theta t)] \tag{129-1}$$

$$\vec{g}_Y(t) = \left(\frac{1}{\cos\theta}\right)^t \cdot [-\vec{g}_X(0) \cdot \sin(\theta t) + \vec{g}_Y(0) \cdot \cos(\theta t)] \tag{129-2}$$

$$\vec{g}_Z(t) = \left(\frac{1}{\cos\theta}\right)^t \cdot \vec{g}_Z(0) \tag{129-3}$$

The path velocity components are observed as:

$$V_X(t) = \left(\frac{1}{\cos\theta}\right)^{2t} \cdot [V_X(0) \cdot \cos(2\theta t) + V_Y(0) \cdot \sin(2\theta t)] \tag{130-1}$$

$$V_Y(t) = \left(\frac{1}{\cos\theta}\right)^{2t} [-V_X(0)\cdot\sin(2\theta t) + V_Y(0)\cdot\cos(2\theta t)] \qquad (130\text{-}2)$$

$$V_Z(t) = \left(\frac{1}{\cos\theta}\right)^{2t} \cdot V_Z(0) \qquad (130\text{-}3)$$

It shows that the original point of Newton inertial coordinator system has a constant absolute velocity $V_X(0), V_Y(0), V_Z(0)$. In fixed position, the path is observed as a spiral curve.

The required $\Theta$ defines the potential field features. According to previous definition, for this kind of motion, the $\dfrac{dx^i}{dt} = f^i(\eta, x)$ in standard rectangular system meets the fowling condition:

$$\frac{\partial f^1}{\partial x^1} = 0, \quad \frac{\partial f^2}{\partial x^2} = 0, \quad \frac{\partial f^3}{\partial x^3} = \frac{1}{\cos\theta} - 1 \qquad (131\text{-}1)$$

$$\frac{\partial f^1}{\partial x^2} = -\frac{\partial f^2}{\partial x^1} = \tan\theta, \text{ others are zero} \qquad (131\text{-}2)$$

For constant spatial feature, $\Theta$ is constant. Its solution of Hamilton dynamic system is:

$$\frac{dx^1}{dt} = f^1(\theta, x) = \tan\theta \cdot x^2 + C_1 \qquad (132\text{-}1)$$

$$\frac{dx^2}{dt} = f^2(\theta, x) = -\tan\theta \cdot x^1 + C_2 \qquad (132\text{-}2)$$

$$\frac{dx^3}{dt} = f^3(\theta, x) = \left(\frac{1}{\cos\theta} - 1\right) \cdot x^3 + C_3 \qquad (132\text{-}3)$$

For $\theta \to \dfrac{\pi}{2}$, (strong potential field), letting $\theta = \dfrac{\pi}{2} - \delta\theta$, as a Newton space approximation, they are simplified as:

$$\frac{dx^1}{dt} = f^1(\delta\theta, x) = \frac{1}{\sin\delta\theta} \cdot x^2 + C_1 \qquad (133\text{-}1)$$

$$\frac{dx^2}{dt} = f^2(\delta\theta, x) = -\frac{1}{\sin\delta\theta} \cdot x^1 + C_2 \qquad (133\text{-}2)$$

$$\frac{dx^3}{dt} = f^3(\delta\theta, x) = \frac{1}{\sin\delta\theta} \cdot x^3 + C_3 \qquad (133\text{-}3)$$

Where, $\delta\theta$ is very small.

Taking the observation point with constant velocity $C_1, C_2, C_3$, in sphere coordinator system, the space continuum is in expansion nearly isotropic. It can be rewritten as:

$$\frac{dR}{dt} = \frac{1}{\sin\delta\theta} \cdot R \qquad (134\text{-}1)$$

Where,

$$\frac{dR}{dt} = \sqrt{\left(\frac{dx^1}{dt} - C_1\right)^2 + \left(\frac{dx^2}{dt} - C_2\right)^2 + \left(\frac{dx^3}{dt} - C_3\right)^2} \qquad (134\text{-}2)$$

$$R = \sqrt{(x^1)^2 + (x^2)^2 + (x^3)^2} \qquad (134\text{-}3)$$

Comparing it with Hubble Law [13],

$$V(R) = \bar{H}_0 \cdot R \tag{135}$$

The Hubble constant $\bar{H}_0$ is determined by the local rotation angular of cosmic spatial continuum as:

$$\bar{H}_0 = \frac{1}{\sin \delta\theta} \tag{136}$$

Hence, Hubble space-time continuum is determined by the singular point of potential field. The Big-bang theory is established based on this $\theta = \frac{\pi}{2}$ singular point.

Around this singular point, the approximated space continuum is approximated as Big-bang theory:

$$\vec{g}_X(t) = \left(\frac{1}{\sin \delta\theta}\right)^t \cdot [\vec{g}_X(0) \cdot \cos(\frac{\pi}{2}t) + \vec{g}_Y(0) \cdot \sin(\frac{\pi}{2}t)] \tag{137-1}$$

$$\vec{g}_Y(t) = \left(\frac{1}{\sin \delta\theta}\right)^t \cdot [-\vec{g}_X(0) \cdot \sin(\frac{\pi}{2}t) + \vec{g}_Y(0) \cdot \cos(\frac{\pi}{2}t)] \tag{137-2}$$

$$\vec{g}_Z(t) = \left(\frac{1}{\sin \delta\theta}\right)^t \cdot \vec{g}_Z(0) \tag{137-3}$$

The path velocity components are observed as:

$$V_X(t) = \left(\frac{1}{\sin \delta\theta}\right)^{2t} \cdot [V_X(0) \cdot \cos(\pi t) + V_Y(0) \cdot \sin(\pi t)] \tag{138-1}$$

$$V_Y(t) = \left(\frac{1}{\sin \delta\theta}\right)^{2t} [-V_X(0) \cdot \sin(\pi t) + V_Y(0) \cdot \cos(\pi t)] \tag{138-2}$$

$$V_Z(t) = \left(\frac{1}{\sin \delta\theta}\right)^{2t} \cdot V_Z(0) \tag{138-3}$$

Based on the physics of finite light velocity as the highest allowable value of mass motion, the minimum time scale (defined by period) $\tilde{T}_{phas}$ is determined by:

$$c = \left(\frac{1}{\sin \delta\theta}\right)^{2\tilde{T}_{phas}} \tag{139}$$

Hence, the minimum time interval can be identified is:

$$\tilde{T}_{phas} = \frac{\ln c}{-\ln(\sin \delta\theta)^2} \tag{140}$$

It is determined by the potential field.

This gives out the uncertainty principle about time as:

$$\Delta t \geq \frac{\ln c}{-\ln(\sin \delta\theta)^2} \tag{141}$$

In one words, the cosmic observation data and the quantum observation features are explained by the theory developed in this research.

*4.7 Uncertainty Principle as the Features of Time-space Continuum*

In time space continuum, by Equation (131) which defines the inertial Newton system, there exists a vacuum Hamilton quantity $H_0$ which is not zero. Observing the equations, it can be rewritten as the conventional form in Hamilton dynamics as:

$$\frac{\partial H_0}{\partial p_1} = \frac{x^2}{R_0} + C_1 \tag{142-1}$$

$$\frac{\partial H_0}{\partial p_2} = -\frac{x^1}{R_0} + C_2 \tag{142-2}$$

$$\frac{\partial H_0}{\partial p_3} = C_3 \tag{142-3}$$

Taking the instant position as reference, letting all $C_i = 0$ (constants be zero), its form solution meets condition:

$$\begin{aligned}dH_0 &= \sqrt{\left(\frac{\partial H_0}{\partial p_1}dp_1\right)^2 + \left(\frac{\partial H_0}{\partial p_1}dp_1\right)^2} \\ &= \frac{1}{R_0}\sqrt{(x^2 dp_1)^2 + (x^1 dp_1)^2} \leq \frac{1}{R_0}\sqrt{[(dp_1)^2 + (dp_1)^2] \cdot [(x^1)^2 + (x^2)^2]}\end{aligned} \tag{143}$$

By suitably selecting time-space unit system, making the item $d\overline{H} = R_0 dH_0$ is the physical energy quantity, as the Plank constant $h$ is the minimum energy measurement, $d\overline{H}_0 \geq h$, while the physical quantity $\Delta p \cdot \Delta x$ has the energy implication, hence one has:

$$\Delta p \cdot \Delta x \geq h = \frac{\hbar}{2\pi} \tag{144}$$

By this way, the uncertainty principle is explained [14] as the intrinsic feature of time-space continuum which is in physical reality as potential field.

It shows that: the uncertainty is the result of curvature of space time for measurement in inertial system.

For the big-bang theory, dark mater, and quantum mechanics, there are many research papers. Here, some papers which give out some similar interpretation as expressed in this research is listed as references [15-19], although I do not accept their theoretic treatment way in form.

## 5. Conclusion

For a mass point motion in potential field described by Hamilton dynamic system, this research that: using geometrical field theory developed in rational mechanics, by defining the spatial continuum configuration deformation produced by the mass motion and defining the acceleration as the Newton velocity transformation, the path velocity transformation of mass motion along its trace are obtained. The stress tensor defined on the mass surface is determined by the spatial continuum physical features and continuum deformation tensor, while the Newton force is determined by the path velocity variation. By this way, the motion equations are obtained.

For constant absolute velocity motion and variation absolute velocity motion, the quantum solutions are obtained. The general principle equations: $\varepsilon = \hbar \cdot \omega$ and $p = \hbar \cdot k$ are obtained. The cosmic expansion model and black hole model are also obtained. The Lorentz parameter, the Hubble parameter and the uncertainty principle are explained also.

The research shows that: the classical physics and the quantum physics are the same in physical intrinsic sense. The only difference is that: the Newton mechanics takes the particle motion as its objective picture, while the quantum mechanics takes the motion path variation as its object picture.

**References**
[1] Y. Nambu, Generalized Hamiltonian Dynamics, Physical Review D, 7:2405-2411, 1973
[2] T. Frankel, The Geometry of Physics, 2$^{nd}$. Cambridge University Press, 2004
[3] V. Kaplunovsky, M. Weinstein, Space-time: Arena or illusion?, Physical Review D, 31(8): 1979-1898, 1985
[4] R. C. Mjolsness, Finite-density nonhomogeneous Newtonian Cosmologies, Phys. Rev. 187:1753-1761, 1968
[5] A. Komar, Hamilton-Jacobi version of general relativity, Phys. Rev. 170:1195-1200, 1968
[6] L. Fernandez-Jambrina, L. M. Gonzalez-Romero, Exterior differential system for cosmological $G_2$ perfect fluids and geodesic completeness, Class. Quantum Grav. 16: 953-972, 1999
[7] R. Bryant, P. Griffiths, D. Grossman, Exterior differential systems and Eular-Lagrange partial differential equations, arXiv: math/0207039v1, [math.dg], 2002
[8] Ulrich H. Gerlach, Derivation of the ten Einstein field equations from the semiclassical approximation to quantum geometrodynamics, Phys. Rev. 177:1929-1941, 1969
[9] B. Carter, Global structure of the Kerr family of gravitational fields, Phys. Rev. 174:1559-1571, 1968
[10] B. S. DeWitt, Quantum theory of gravity.1. the canonical theory, Phys. Rev. 160:1113-1148, 1967
[11] R. Arnowitt, Dynamical structure and definition of energy in general relativity, Phys. Rev. 116:1322-1330, 1959
[12] Xiao Jianhua. Evolution of Continuum from Elastic Deformation to Flow. E-print, arXiv: physics/0511170(physics.class-ph), 2005
[13] He Xiantao. Observational Cosmics. Beijing: Science Pub., 2002 (In Chinese)
[14] Xiao Jianhua, On the Rational Relationship between Heisenberg Principle and General Relativity, arXiv: Physics/0610237(physics.gen-ph), 2006
[15] C. L. Herzberg, Mass and the creation of spatial volume, arXiv:physics/1105.2705 (physics.gen-ph), 2011
[16] C. L. Herzberg, Holographic position uncertainty and the quantum-classical transition, arXiv:physics/1004.2642, (phyics.gen-ph), 2010
[17] C. L. Herzberg, Can the expansion of the universe localize quantum wave functions? Howclassical behavior may result from Hubble expansion, arXiv:physics/0912.1158, (physics.gen-ph), 2009
[18] C. L. Herzberg, Dark matter and the origin of mass: a simple semiclassical approach, arXiv:physics/0006040, (physics.gen-ph), 2000
[19] V. N. Borodikhin, Explanation of velocities distribution in the galaxies without the dark matter, arXiv:phyics/1105.1709 (physics.gen-ph), 2011